\documentclass{emulateapj}

\def\kms{\ifmmode{\rm km\thinspace s^{-1}}\else km\thinspace s$^{-1}$\fi}
\def\vstar{V541\,Cyg}

\shortauthors{Torres et al.}
\shorttitle{\vstar}

\begin{document}

\submitted{Accepted for publication in The Astrophysical Journal}

\title{Absolute dimensions of the eccentric eclipsing binary V541 Cygni}

\author{
Guillermo Torres\altaffilmark{1},
Chima D.\ McGruder\altaffilmark{1,2},
Robert J.\ Siverd\altaffilmark{3},
Joseph E.\ Rodriguez\altaffilmark{1},
Joshua Pepper\altaffilmark{4},
Daniel J.\ Stevens\altaffilmark{5},
Keivan G. Stassun\altaffilmark{6,7},
Michael B.\ Lund\altaffilmark{6}, and
David James\altaffilmark{8}
}

\altaffiltext{1}{Harvard-Smithsonian Center for Astrophysics, 60
  Garden St., Cambridge, MA 02138, USA; e-mail:
  gtorres@cfa.harvard.edu}

\altaffiltext{2}{University of Tennessee, Knoxville, 1408 Circle
  Drive, Knoxville, TN 37996, USA}

\altaffiltext{3}{Las Cumbres Observatory Global Telescope Network, 6740
  Cortona Dr., Suite 102, Santa Barbara, CA 93117, USA}

\altaffiltext{4}{Lehigh University, Department of Physics, 413 Deming
  Lewis Lab, 16 Memorial Drive East Bethlehem, PA 18015, USA}

\altaffiltext{5}{Department of Astronomy, The Ohio State University,
  Columbus, OH 43210, USA}

\altaffiltext{6}{Department of Physics and Astronomy, Vanderbilt
  University, 6301 Stevenson Center, Nashville, TN 37235, USA}

\altaffiltext{7}{Department of Physics, Fisk University, 1000 17th
  Avenue North, Nashville, TN 37208, USA}

\altaffiltext{8}{Astronomy Department, University of Washington, Box
  351580, Seattle, WA 98195, USA}

\begin{abstract}

We report new spectroscopic and photometric observations of the
main-sequence, detached, eccentric, double-lined eclipsing binary
\vstar\ ($P = 15.34$~days, $e = 0.468$). Using these observations
together with existing measurements we determine the component masses
and radii to better than 1\% precision: $M_1 =
2.335^{+0.017}_{-0.013}~M_{\sun}$, $M_2 =
2.260^{+0.016}_{-0.013}~M_{\sun}$, $R_1 =
1.859^{+0.012}_{-0.009}~R_{\sun}$, and $R_2 =
1.808^{+0.015}_{-0.013}~R_{\sun}$. The nearly identical B9.5 stars
have estimated temperatures of $10650 \pm 200$~K and $10350 \pm
200$~K. A comparison of these properties with current stellar
evolution models shows excellent agreement at an age of about 190~Myr
and ${\rm [Fe/H]} \approx -0.18$. Both components are found to be
rotating at the pseudo-synchronous rate. The system displays a slow
periastron advance that is dominated by General Relativity (GR), and
has previously been claimed to be slower than predicted by theory. Our
new measurement, $\dot\omega =
0.859^{+0.042}_{-0.017}$~deg~century$^{-1}$, has an 88\% contribution
from GR and agrees with the expected rate within the uncertainties. We
also clarify the use of the gravity darkening coefficients in the
light-curve fitting program EBOP, a version of which we use here.

\end{abstract}

\keywords{ binaries: eclipsing --- stars: evolution --- stars:
  fundamental parameters --- stars: individual (\vstar) ---
  techniques: photometric }

%%%%%%%%%%%%%%%%%%%%%%%%%%%%%%%%%%%%%%%%%%%%%%%%%%%%%%%%%%%%%%%%%%%%%%%%%%%
\section{Introduction}
\label{sec:introduction}
%%%%%%%%%%%%%%%%%%%%%%%%%%%%%%%%%%%%%%%%%%%%%%%%%%%%%%%%%%%%%%%%%%%%%%%%%%%

\vstar\ (also HD~331102, BD+30~3704, TYC~2656-3703-1, $V = 10.44$) is
an early type (B9.5+B9.5) detached, eccentric, double-lined eclipsing
binary with a relatively long period of 15.33~days
\citep{Kulikowski:1948, Kulikowski:1953} and nearly identical
components. The first photoelectric light curve was obtained and
analyzed by \cite{Khaliullin:1985}, but the masses could not be
determined dynamically because no spectroscopic observations were
available at the time. This was remedied by \cite{Lacy:1998}, who
reported radial-velocity measurements for both components and analyzed
them in conjunction with the $V$-band light curve from
\cite{Khaliullin:1985} to obtain the absolute masses
($\approx$2.2~$M_{\sun}$) and radii ($\approx$1.8~$R_{\sun}$) with
relative errors of about 4\% and 2\%, respectively.

The system is noteworthy in that it presents a periastron advance that
is dominated by the general relativistic effect, estimated to be
several times larger than the classical effects due to tidal and
rotational distortions. However, there has been some disagreement over
the precise rate of apsidal motion, which is fairly slow and difficult
to determine, and how closely it conforms to theoretical expectations.
Some authors have obtained good agreement with the predicted motion,
while others have measured a rate of precession that is too slow, and
argued that \vstar\ may belong to a small group of binaries including
DI~Her and AS~Cam that display similar discrepancies. Past
speculations about possible shortcomings of General Relativity or
alternative theories of gravitation that might explain the puzzle
\citep[e.g.,][]{Guinan:1985} have largely fallen out of favor, and at
least in the case of DI~Her the ``anomalous'' apsidal motion has now
been shown to be caused by the tilted spin axes of the stars relative
to the axis of the orbit \citep{Company:1988, Albrecht:2009}.

The motivation for this paper is twofold. Firstly, we note that the
precision of the mass and radius estimates by \cite{Lacy:1998} was
limited by the quality and quantity of his spectroscopic observations.
In particular, the masses are not quite precise enough for a
meaningful comparison with modern stellar evolution models for this
important binary system \citep[see, e.g.,][]{Torresetal:2010}, a test
that was not performed in the original study by that author. To this
end we have obtained a much richer set of high-resolution spectra that
allows substantial improvement in the absolute dimensions of the two
stars.  We also bring to bear additional light curves obtained more
recently to supplement the only existing set of photometric
measurements by \cite{Khaliullin:1985}.  Secondly, we wish to revisit
the determination of the apsidal motion and the comparison with
internal structure theory by taking advantage of all existing
measurements (light curves, radial-velocity measurements, eclipse
timings) in a self-consistent way. Previously this has been done using
only the measured times of eclipse.

The paper is organized as follows. Section~\ref{sec:spectroscopy} and
Section~\ref{sec:photometry} describe our new spectroscopic and
photometric observations of \vstar, and the times of minimum light
that have the longest time coverage are presented in
Section~\ref{sec:timings}. Our combined analysis of all the data is
described in Section~\ref{sec:analysis}, followed by a summary of the
inferred physical properties of the components in
Section~\ref{sec:dimensions}. Then in
Section~\ref{sec:evolution} we discuss the comparison of the
mass, radius, and temperature determinations for \vstar\ against
current models of stellar evolution, and our new measurement of the
rate of apsidal motion of the binary is presented in
Section~\ref{sec:apsidal}. We end with a discussion of these results
and final remarks in Section~\ref{sec:discussion}.

%%%%%%%%%%%%%%%%%%%%%%%%%%%%%%%%%%%%%%%%%%%%%%%%%%%%%%%%%%%%%%%%%%%%%%%%%%%
\section{Spectroscopic observations}
\label{sec:spectroscopy}
%%%%%%%%%%%%%%%%%%%%%%%%%%%%%%%%%%%%%%%%%%%%%%%%%%%%%%%%%%%%%%%%%%%%%%%%%%%

\vstar\ was observed spectroscopically at the Harvard-Smithsonian
Center for Astrophysics (CfA) with the Digital Speedometer
\citep[DS;][]{Latham:1992} on the 1.5m Tillinghast reflector at the
Fred L.\ Whipple Observatory on Mount Hopkins (AZ). This instrument
(now decommissioned) was an echelle spectrograph with a resolving
power of $R \approx 35,000$ equipped with a photon-counting
intensified Reticon detector limiting the output to a single order
45~\AA\ wide centered on the \ion{Mg}{1}\,b triplet at 5187~\AA.  We
gathered 72 exposures between 2000 July and 2004 October with
signal-to-noise ratios ranging from 20 to 49 per resolution element of
8.5~\kms. One of the observations was obtained during an eclipse and
was excluded from further analysis. Reductions were performed with a
custom pipeline, and the wavelength calibration was based on exposures
of a Thorium-Argon lamp before and after each science exposure.
Exposures of the dusk and dawn sky were taken regularly for the
purpose of monitoring instrumental drifts. All spectra are clearly
double-lined.

Radial velocities (RVs) were determined by cross-correlation using the
two-dimensional algorithm TODCOR \citep{Zucker:1994}. Templates (one
for each component) were taken from a large library of synthetic
spectra based on model atmospheres by R.\ L.\ Kurucz
\citep[see][]{Nordstrom:1994, Latham:2002}, computed for a range of
temperatures ($T_{\rm eff}$), surface gravities ($\log g$), rotational
broadenings ($v \sin i$ when seen in projection), and metallicities
([m/H]). The optimum templates were selected by cross-correlating each
of our 71 spectra against synthetic spectra covering a wide range of
parameters, and seeking the best match as measured by the maximum
cross-correlation coefficient averaged over all exposures and weighted
by the strength of each spectrum \citep[see][]{Torres:2002}.

Experience has shown that the narrow wavelength coverage of our
spectra introduces strong correlations between $T_{\rm eff}$, $\log
g$, and [m/H] such that optimal matches of similar quality can be
obtained by slightly increasing or decreasing these three parameters
in tandem. We therefore assumed initially that the metallicity is
solar, and held $\log g$ fixed at values near those determined later
in our analysis (Section~\ref{sec:dimensions}). Because the grid
spacing of our library in $\log g$ is 0.5~dex and the actual values
($\sim$4.25 for both stars) are intermediate between two grid points,
we repeated the determinations using $\log g$ values of 4.0 and 4.5,
and interpolated the results. Similarly, the comparison with stellar
evolution models described in Section~\ref{sec:evolution} points to a
composition intermediate between ${\rm [m/H]} = 0.0$ and $-0.5$, so we
repeated the determinations at the lower metallicity (for each of the
two values of $\log g$) and again interpolated. The resulting best-fit
temperatures of the slightly larger and more massive primary component
(star~1) and of the secondary (star~2) are 10650~K and 10350~K,
respectively, with estimated uncertainties of 200~K. The $v \sin i$
values are $15 \pm 1~\kms$ for both stars.  Templates near these
best-fit values were used to derive the radial velocities. The flux
ratio between the secondary and primary as determined with TODCOR is
$\ell_2/\ell_1 = 0.92 \pm 0.02$ at the mean wavelength of our
observations, 5187~\AA.

Systematic errors in the radial velocities that may result from
spectral lines shifting in and out of the narrow spectral window as a
function of orbital phase were investigated by means of numerical
simulations as described by \cite{Torres:1997}. Briefly, we generated
artificial composite spectra matching each of the real spectra by
combining the two templates scaled by the above flux ratio and with
the proper relative Doppler shifts as determined from a preliminary
spectroscopic orbital solution. We then processed these artificial
spectra with TODCOR in the same way as the real spectra, and compared
the input and output Doppler shifts. The differences were applied as
corrections to the raw velocities measured from the real spectra, and
were typically smaller than 0.5~\kms, which is only about 1/3 of our
internal errors. The final radial velocities in the heliocentric frame
are listed in Table~\ref{tab:rvs} and include these corrections for
systematics as well as small run-to-run adjustments to remove
instrumental drifts, as described above. Typical uncertainties are
about 1.5~\kms.

\begin{deluxetable*}{lccccccc}
\tablewidth{0pc}
\tablecaption{Heliocentric radial velocity measurements of \vstar\ from CfA.
 \label{tab:rvs}}
\tablehead{
\colhead{HJD} &
\colhead{$RV_1$} &
\colhead{$\sigma_1$} &
\colhead{$(O-C)_1$} &
\colhead{$RV_2$} &
\colhead{$\sigma_2$} &
\colhead{$(O-C)_2$} &
\colhead{Orbital}
\\
\colhead{(2,400,000$+$)} &
\colhead{(\kms)} &
\colhead{(\kms)} &
\colhead{(\kms)} &
\colhead{(\kms)} &
\colhead{(\kms)} &
\colhead{(\kms)} &
\colhead{phase}
}
\startdata
51740.7766  &  $-$57.26  &   1.35  &  $-$0.27  &   27.09  &   1.30  &   $-$0.59  &  0.1654 \\ 
51743.8481  &  $-$96.49  &   2.72  &  $+$2.75  &   69.96  &   2.62  &   $-$1.32  &  0.3657 \\ 
51743.8753  &  $-$98.75  &   1.62  &  $+$0.34  &   69.06  &   1.56  &   $-$2.07  &  0.3675 \\ 
51800.7427  &  $-$37.09  &   1.31  &  $-$0.95  &    4.91  &   1.26  &   $-$1.26  &  0.0751 \\ 
51802.6632  &  $-$64.62  &   1.39  &  $+$1.02  &   36.69  &   1.34  &   $+$0.09  &  0.2003 
\enddata
\tablecomments{This table is available in its entirety in
  machine-readable form.}
\end{deluxetable*}

In addition to our own observations we have made use in our analysis
below of the 16 radial-velocity measurements by \cite{Lacy:1998},
which, although less numerous and of lower precision than our own,
were obtained much earlier (1982--1994) and can help to constrain the
periastron advance.

%%%%%%%%%%%%%%%%%%%%%%%%%%%%%%%%%%%%%%%%%%%%%%%%%%%%%%%%%%%%%%%%%%%%%%%%%%%
\section{Photometric observations}
\label{sec:photometry}
%%%%%%%%%%%%%%%%%%%%%%%%%%%%%%%%%%%%%%%%%%%%%%%%%%%%%%%%%%%%%%%%%%%%%%%%%%%

The $V$-band observations of \cite{Khaliullin:1985}, which we
reanalyze here, consist of 531 differential measurements obtained
between 1981 May and 1983 July with a 0.5m reflector at the Crimea
Observatory of the Sternberg Astronomical Institute. The comparison
star was BD+30~3702 (HD~331103) and the check star was BD+31~3728
(HD~331101).  The reported internal precision of these measurements is
0.009~mag.

More recent observations of \vstar\ were obtained between 2007 May and
2014 November in the course of the Kilodegree Extremely Little
Telescope transiting planet program \citep[KELT;][]{Pepper:2007}. KELT
is an all-sky photometric survey to discover transiting planets around
bright host stars ($7 < V < 11$). It uses two telescopes: KELT-North
located in Sonoita (AZ), and KELT-South at the South African
Astronomical Observatory (SAAO). Both telescope systems are based on a
Mamiya 645-series wide-angle 42mm lens with an 80mm focal length
($f$/1.9) giving a field of view of $26\arcdeg \times 26\arcdeg$. The
observations of \vstar\ reported here were gathered at the northern
site with a $4096 \times 4096$ pixel Apogee AP16E CCD camera (9~$\mu$m
pixels, corresponding to 23\arcsec\ on the sky) and a 10--20 minute
cadence.  \vstar\ is located in KELT-North field 11 ($\alpha$ =
19$^{\rm h}$\,27$^{\rm m}$\,00$^{\rm s}$, $\delta$ =
$31\arcdeg$\,39$\arcmin$\,56\arcsec, J2000). The telescopes are
mounted on a Paramount ME German equatorial mount causing images taken
East of the meridian to have a 180\arcdeg\ rotation compared to the
images to the West \citep{Pepper:2007, Pepper:2012}.  As the optics
for each telescope are not perfectly axisymmetric, the point spread
function (PSF) of the same star in the corner of the images is not the
same in each orientation. In addition, the PSF
full-width-at-half-maximum (FWHM) varies between 3 and 6 pixels
depending on where the star is positioned on the CCD. Therefore, the
East and West images have been treated as independent time series
throughout the reduction and detrending process. Because of the
large PSF ($\sim$1--2\arcmin\ on the sky), light from neighboring
stars is likely to affect the photometry and must be taken to account,
as described in Section~\ref{sec:analysis}. The passband of these
observations resembles that of a very broad $R$-band filter.

Reductions and detrending were carried out with a dedicated pipeline
that produces three types of light curves for each target star in the
program: a raw extracted light curve, a ``scaled'' version that is the
raw light curve with a 90-day median smoothing applied, and a detrended
version that uses the Trend Filtering Algorithm
\citep[TFA;][]{Kovacs:2005}. For a detailed description of the KELT
data acquisition, reduction, and post-processing, see
\cite{Siverd:2012} and \cite{Kuhn:2016}. Because the TFA procedures
are tuned for the detection of very shallow, transit-like signals
characteristic of exoplanets, the results for objects with very deep
($\sim$0.7~mag) eclipses such as those in \vstar\ are less than
optimal. Therefore, for the analysis in this paper we have made use of
the ``scaled'' light curves. The time series East of the meridian
consists of 3885 measurements with a typical internal photometric precision of
0.017~mag, and the one to the West has 2940 measurements with typical
uncertainties of 0.013~mag. These observations are reported in
Table~\ref{tab:KELTeast} and Table~\ref{tab:KELTwest}, respectively.

\begin{deluxetable}{lcc}
\tablewidth{0pc}
\tablecaption{KELT observations of \vstar\ East of the meridian.
 \label{tab:KELTeast}}
\tablehead{
\colhead{HJD-2,400,000} &
\colhead{$R$ (mag)} &
\colhead{$\sigma$ (mag)}
}
\startdata
54257.770944  &  14.941    &  0.016   \\
54257.775454  &  14.926    &  0.015   \\
54257.779964  &  14.903    &  0.015   \\
54257.784484  &  14.934    &  0.015   \\
54257.788995  &  14.915    &  0.016  
\enddata
\tablecomments{This table is available in its entirety in
  machine-readable form.}
\end{deluxetable}

\begin{deluxetable}{lcc}
\tablewidth{0pc}
\tablecaption{KELT observations of \vstar\ West of the meridian.
 \label{tab:KELTwest}}
\tablehead{
\colhead{HJD-2,400,000} &
\colhead{$R$ (mag)} &
\colhead{$\sigma$ (mag)}
}
\startdata
54250.936923  &  14.615    &  0.013   \\
54250.942503  &  14.637    &  0.013   \\
54276.867035  &  14.658    &  0.011   \\
54362.631612  &  14.637    &  0.011   \\
54363.630845  &  14.665    &  0.012  
\enddata
\tablecomments{This table is available in its entirety in
  machine-readable form.}
\end{deluxetable}

%%%%%%%%%%%%%%%%%%%%%%%%%%%%%%%%%%%%%%%%%%%%%%%%%%%%%%%%%%%%%%%%%%%%%%%%%%%
\section{Times of minimum light}
\label{sec:timings}
%%%%%%%%%%%%%%%%%%%%%%%%%%%%%%%%%%%%%%%%%%%%%%%%%%%%%%%%%%%%%%%%%%%%%%%%%%%

Numerous times of minimum light have been recorded for \vstar\ and
used over the years to improve the ephemeris.  The more recent CCD
measurements carry essentially all of the weight, and older
photographic or visual timings that have occasionally been included
have such large scatter that they are of little use for determining
the apsidal motion.  Prior to the publication of the photoelectric
light curve of \cite{Khaliullin:1985} the most complete light curve of
\vstar\ was the one reported by \cite{Karpowicz:1961}, based on his
extensive series of 161 photographic plates collected with a 14\,cm
astrograph at Ostrowik near Warsaw (Poland) between 1955 and 1959.
From these observations Karpowicz reported three average times of
minimum light for the primary and three for the secondary, each
determined from plates restricted to time spans of no more than two
years. Though perhaps better than other photographic estimates, even
those timings give very large residuals by modern standards.

However, it is still possible to extract useful information from this
material in the form of one average time of primary minimum and one
average time of secondary minimum for the entire series. This can be
done by using all of Karpowicz' original photographic measurements
together, which allows more complete coverage of the eclipses and
therefore a better definition of the minima. The apsidal motion is
slow enough that any change in the separation between primary and
secondary eclipse over the four-year interval of these observations is
small compared to the uncertainties. We proceeded as follows. We
phase-folded the 161 brightness measurements with a preliminary
ephemeris based on the analysis below, and then fit the primary and
secondary eclipses separately using as a template a model light curve
from a preliminary solution to the $V$-band data of
\cite{Khaliullin:1985}.  After applying a vertical offset to the
template by eye, we allowed only a horizontal (phase) shift to fit
each eclipse (keeping the shape fixed), and finally converted the
phase shifts back to time units. We assigned these measurements to a
reference epoch closest to the average time of all photographic
observations, obtaining ${\rm Min~I} = 2,\!436,\!262.3278 \pm 0.0031$
(HJD) and ${\rm Min~II} = 2,\!436,\!269.3158 \pm 0.0031$ (HJD).
Table~\ref{tab:timings} lists these two timings along with all other
published photoelectric/CCD timings. We note that, although the table
also lists the two timings reported by \cite{Khaliullin:1985}, we do
not use them here because our analysis incorporates the full $V$-band
light curve from this author, which contains the same information.

\begin{deluxetable}{ll@{}c@{}cc@{}c@{}}
%\tablewidth{0pc}
\tablecaption{Times of minimum for \vstar.\label{tab:timings}}
\tablehead{
\colhead{HJD} &
\colhead{$\sigma$} &
\colhead{} &
\colhead{$(O-C)$} &
\colhead{} &
\colhead{} \\
\colhead{(2,400,000$+$)} &
\colhead{(days)} &
\colhead{Type} &
\colhead{(days)} &
\colhead{Year} &
\colhead{Source}}
\startdata
36262.3278                  &   0.0031  &  1  &  $-$0.00007  &  1958.1583  &  1  \\
36269.3158                  &   0.0031  &  2  &  $+$0.00151  &  1958.1775  &  1  \\
44882.2148\tablenotemark{a} &   0.0007  &  1  &  $-$0.00026  &  1981.7583  &  2  \\
44889.2196\tablenotemark{a} &   0.0005  &  2  &  $-$0.00006  &  1981.7775  &  2  \\
46998.8424                  &   0.0010  &  1  &  $+$0.00024  &  1987.5533  &  3  \\
48616.3400\tablenotemark{b} &  \nodata  &  2  &  $+$0.00823  &  1991.9818  &  4  \\
48839.3870                  &   0.003   &  1  &  $-$0.00046  &  1992.5924  &  5  \\
49168.4947                  &   0.0007  &  2  &  $-$0.00183  &  1993.4935  &  6  \\
49168.4955                  &   0.0009  &  2  &  $-$0.00103  &  1993.4935  &  6  \\
49560.2668                  &   0.0008  &  1  &  $-$0.00091  &  1994.5661  &  7  \\
49889.3770                  &   0.001   &  2  &  $-$0.00130  &  1995.4672  &  8  \\
49904.7145                  &   0.00012 &  2  &  $-$0.00171  &  1995.5091  &  9  \\
49935.3911                  &   0.0006  &  2  &  $-$0.00093  &  1995.5931  & 10  \\
49935.3974\tablenotemark{b} &   0.0012  &  2  &  $+$0.00537  &  1995.5931  & 11  \\
50296.4906                  &   0.0018  &  1  &  $+$0.00477  &  1996.5818  & 12  \\
51070.3967                  &  \nodata  &  2  &  $-$0.00066  &  1998.7006  & 13  \\
51109.3918                  &  \nodata  &  1  &  $-$0.00154  &  1998.8074  & 13  \\
51385.4740                  &   0.0007  &  1  &  $-$0.00113  &  1999.5632  & 14  \\
52926.28338                 &   0.0001  &  2  &  $-$0.00108  &  2003.7817  & 15  \\
53524.46196                 &   0.0001  &  2  &  $-$0.00099  &  2005.4195  & 15  \\
53578.7911                  &   0.0001  &  1  &  $-$0.00052  &  2005.5682  & 16  \\
53846.55486\tablenotemark{b}&   0.0007  &  2  &  $-$0.00420  &  2006.3013  & 15  \\
54659.46742                 &   0.0002  &  2  &  $-$0.00086  &  2008.5269  & 17  \\
54659.46762                 &   0.0002  &  2  &  $-$0.00066  &  2008.5269  & 17  \\
54659.46809                 &   0.0001  &  2  &  $-$0.00019  &  2008.5269  & 15  \\
54782.1712                  &   0.0003  &  2  &  $-$0.00036  &  2008.8629  & 15  \\
55066.5740                  &   0.0024  &  1  &  $+$0.00827  &  2009.6415  & 18  \\
55741.42898                 &   0.0007  &  1  &  $-$0.00336  &  2011.4892  & 19  \\
55741.4299                  &   0.0064  &  1  &  $-$0.00244  &  2011.4892  & 20  \\
55741.43011                 &   0.0010  &  1  &  $-$0.00223  &  2011.4892  & 19  \\
55794.4734                  &   0.0017  &  2  &  $-$0.00022  &  2011.6344  & 21  \\
56876.4377                  &   0.0026  &  1  &  $+$0.00242  &  2014.5967  & 22  \\
56929.4809                  &   0.0020  &  2  &  $+$0.00195  &  2014.7419  & 23  \\
57198.5271                  &   0.0026  &  1  &  $-$0.00361  &  2015.4785  & 24 
\enddata
\tablecomments{Measurement errors ($\sigma$) are listed as
  published. ``Type'' is 1 for a primary eclipse, 2 for a secondary
  eclipse. Uncertainties for the timings with no published errors are
  assumed to be 0.001~days.  $O\!-\!C$ residuals are computed from the
  combined fit described in Section~\ref{sec:analysis}.  Sources for
  the times of minimum light are:
(1) Measurement made in this paper (see text) based on the
    photographic light curve of \cite{Karpowicz:1961};
(2) \cite{Volkov:1999};
(3) \cite{Lines:1989};
(4) \cite{Diethelm:1992a};
(5) \cite{Diethelm:1992b};
(6) \cite{Agerer:1994};
(7) \cite{SandbergLacy:1995};
(8) \cite{Wolf:1995};
(9) \cite{Guinan:1996};
(10) \cite{Lacyetal:1998};
(11) \cite{Diethelm:1995};
(12) \cite{Diethelm:1996};
(13) \cite{Volkov:1999};
(14) \cite{Diethelm:1999};
(15) \cite{Wolf:2010};
(16) \cite{Smith:2007};
(17) \cite{Brat:2008};
(18) \cite{Hubscher:2010};
(19) \cite{Honkova:2013};
(20) \cite{Hubscheretal:2012};
(21) \cite{Hubscher:2012};
(22) \cite{HubscherLehmann:2015};
(23) \cite{Hubscher:2015};
(24) \cite{Hubscher:2016}.
}
\tablenotetext{a}{Timing from \cite{Khaliullin:1985} revised by
\cite{Volkov:1999} and ignored in the fit of Section~\ref{sec:analysis}
because the same information is already contained in the original
light curve used in the global fit.}
\tablenotetext{b}{Measurement not used in the global fit for showing a
  very large residual ($> 4\sigma$).}
\end{deluxetable}

In general one expects the times of secondary minimum to be more
precisely determined than those of primary minimum for \vstar, on
account of the much shorter duration of the secondary eclipse
(6~hours) compared to the primary (16~hours) caused by the large
eccentricity and orientation of the orbit. However, this does not seem
to be reflected in the formal uncertainties shown in
Table~\ref{tab:timings}, possibly due to the heterogeneous nature of
these measurements. Many studies have shown that formal timing errors
tend to be underestimated, and our analysis below supports this.

%%%%%%%%%%%%%%%%%%%%%%%%%%%%%%%%%%%%%%%%%%%%%%%%%%%%%%%%%%%%%%%%%%%%%%%%%%%
\section{Analysis}
\label{sec:analysis}
%%%%%%%%%%%%%%%%%%%%%%%%%%%%%%%%%%%%%%%%%%%%%%%%%%%%%%%%%%%%%%%%%%%%%%%%%%%

\begin{figure}
\epsscale{1.15}
\plotone{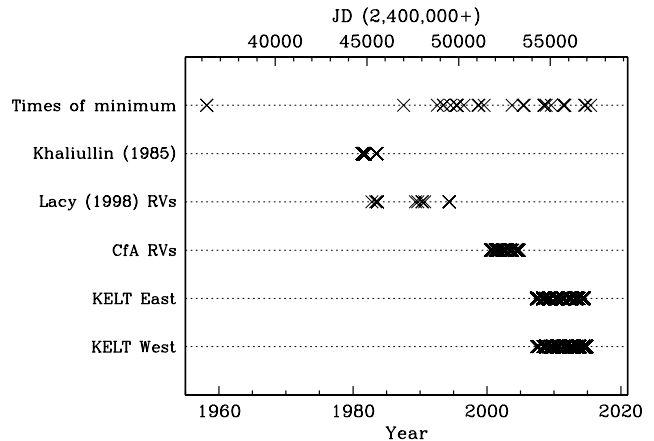}
\figcaption[]{Time history of the photometric and spectroscopic
  observations of \vstar\ used in this paper.\label{fig:history}}
\end{figure}

The various data sets available for \vstar\ are complementary in many
ways. For example, the light curve of \cite{Khaliullin:1985} is of
high quality but has incomplete coverage of the primary eclipse. The
KELT data, on the other hand, have somewhat lower precision but offer
essentially complete coverage of all phases, aiding in determining
some of the geometric properties of the orbit. These two photometric
time series provide strong constraints on the location of the line of
apsides at two epochs some three decades apart. The times of minimum
light from the previous section provide additional constraints on the
apsidal motion over a longer time span of 57 years (see
Figure~\ref{fig:history}). Finally the radial velocities yield
constraints of a different nature at other times, particularly on the
eccentricity of the orbit, so that the optimum approach is to combine
all of these data in a global fit to solve for all orbital parameters
simultaneously, including the apsidal motion.

For the analysis of the light curves of this well-detached system we
have adopted the Nelson-Davis-Etzel model \citep{Popper:1981,
  Etzel:1981}, implemented originally in the widely used EBOP code.
The particular version of the light-curve generator in EBOP used here
is a rewrite due to \cite{Irwin:2011} that also handles apsidal
motion\footnote{\url https://github.com/mdwarfgeek/eb~.}, and is
especially useful within the framework of the Markov Chain Monte Carlo
methodology we apply here. The parameters solved for include the
sidereal period $P_{\rm sid}$, a reference time of primary eclipse
$T_0$ chosen to be near the mean of all our times of observation to
minimize correlations, the sum of the relative radii $r_1+r_2$
normalized to the semimajor axis, the radius ratio $k \equiv r_2/r_1$,
the cosine of the inclination angle $\cos i$, the eccentricity
parameters $\sqrt{e}\cos\omega$ and $\sqrt{e}\sin\omega$ at $T_0$, the
rate of apsidal motion $\dot\omega$, the central surface brightness
ratio $J \equiv J_2/J_1$, the fraction of third light divided by the
total light ($L_3$), and a magnitude zero point $m_0$.  Light travel
time across the system was accounted for, although the effect is
hardly noticeable.

For the limb darkening effect we adopted the linear law with a
coefficient $u$ (assumed to be the same for the two nearly identical
components), as experiments with higher-order prescriptions indicated
no improvement. The gravity darkening input required by the code is
the product of two values: the exponent $\beta$ of the law relating
the emergent bolometric flux to the local surface gravity, and a
separate wavelength-specific gravity darkening coefficient,
$y_{\lambda}$.\footnote{There has been some confusion in the
  literature regarding the nature of the input values for gravity
  darkening expected by the EBOP program in its various
  implementations, beginning with the original version
  \citep{Popper:1981, Etzel:1981} and continuing with its descendants
  such as JKTEBOP \citep{Southworth:2013} and the code used here
  \citep{Irwin:2011}.  While other light-curve fitting programs deal
  internally with the wavelength dependence of the flux $F$ as
  affected by local gravity $g$ using either the blackbody formula or
  model atmospheres, and therefore expect the input values for gravity
  darkening to be the exponent ($\beta$ in recent usage) of the
  bolometric law $F \propto g^\beta$ for each star
  \citep{vonZeipel:1924, Lucy:1967, Claret:1998}, EBOP has no
  knowledge of the passband of the observations, or in fact of the
  absolute temperatures of the stars (by design).  Therefore, it is up
  to the user to supply values appropriate for the wavelength
  $\lambda$ and stellar temperatures, as is the case for the
  limb-darkening coefficients. EBOP adopts a simple Taylor expansion
  of the wavelength-dependent flux as a function of local gravity of
  the form $F_{\lambda} = F_{0,\lambda} \left[1 + y_{\lambda} \beta
    (g-g_0)/g_0\right]$ \citep[see, e.g.,][]{Kopal:1959,
    Binnendijk:1960, Martynov:1973, Kitamura:1983}, retaining only the
  linear term consistent with the level of other approximations and
  the intended application of the model to simple, well-detached
  binary systems.  The input required by EBOP is the {\it product} of
  $\beta$ and the gravity darkening coefficient $y_{\lambda}$, which
  depends on wavelength and also temperature. Expressions to compute
  $y_{\lambda}$ are provided in the above references using a blackbody
  approximation, which is sufficient for our purposes given that the
  gravity darkening effect is essentially negligible for nearly
  spherical stars such as those in \vstar. Note that $y$ is unity for
  the total (bolometric) flux, but varies significantly as a function
  of $T_{\rm eff}$ and $\lambda$ and can be larger or smaller than
  unity. Although the prescription for gravity darkening in terms of
  the coefficient $y$ is described in the original documentation for
  EBOP \citep{Etzel:1980}, where the implicit assumption is also made
  that $\beta$ is always unity, that document is unpublished and has
  generally been difficult to obtain, which may have contributed to
  the confusion.} Here we assume $\beta = 1$, as appropriate for
radiative stars, and the coefficients $y_{\lambda}$ were set for each
passband to values appropriate for the stellar temperatures: 0.67 and
0.68 for the primary and secondary in $V$, and 0.59 and 0.61 for the
KELT passband. Separate values of the wavelength-dependent parameters
$J$ and $u$ were considered for the \cite{Khaliullin:1985} data and
for the two KELT light curves, and magnitude zero points were solved
independently for each of the three light curves. Inclusion of the
third light parameter was motivated in part by the report of the
presence of a close visual companion to \vstar\ in the {\it Tycho}
Double Star Catalogue \citep{Fabricius:2002}, at an angular separation
of 0\farcs75 in position angle 6\fdg2 and with a reported brightness
difference of about 1.2~mag in the $V_T$ passband. We note, however,
that subsequent observations by \cite{Mason:2009} using speckle
interferometry did not detect this companion, even though they were
sensitive to it. One value of $L_3$ was included for the $V$-band
data, and two additional values were used for the KELT data sets
because their much larger PSF ($\sim$1--2\arcmin;
Section~\ref{sec:photometry}) is different for the east and west (see
Section~\ref{sec:photometry}).

For the radial velocities we assumed pure Keplerian orbital motion,
and solved for the velocity semi-amplitudes $K_1$ and $K_2$, the
center-of-mass velocity $\gamma$, as well as for a possible offset
$\Delta$ between the velocity zero points for the primary and
secondary. The latter may arise, for example, from template mismatch
in the cross-correlation procedure. The parameters $\gamma$ and
$\Delta$ were solved independently for our own RVs and those of
\cite{Lacy:1998}.

The times of minimum light were incorporated using the formalism of
\cite{Lacy:1992} for the ephemeris curve solution. We note that the
EBOP light-curve generator used here assumes $T_0$ is a time of
inferior conjunction, rather than a time of minimum light, but the
difference for this system is negligible.

Our method of solution used the {\tt emcee\/}\footnote{\url
  http://dan.iel.fm/emcee~.} code of \cite{Foreman-Mackey:2013}, which
is a Python implementation of the affine-invariant Markov Chain Monte
Carlo ensemble sampler proposed by \cite{Goodman:2010}. Uniform priors
over suitable ranges were used for most parameters, and modified
Jeffreys priors were used for third light and the jitter parameters
\citep{Gregory:2005}, although the results are insensitive to these
assumptions.  Relative weighting between the different data sets was
handled by including additional adjustable parameters to inflate the
observational errors, which were solved for self-consistently and
simultaneously with the other parameters \citep[see][]{Gregory:2005}.
To inflate the uncertainties of the three light curves we adopted
simple scale factors $f(V)$, $f({\rm KELT~east})$, and $f({\rm
  KELT~west})$, as is customary, and for the radial velocities and the
times of minimum we used separate ``jitter'' terms for the primary and
secondary ($\epsilon_{\rm RV,1}$, $\epsilon_{\rm RV,2}$,
$\epsilon_{\rm Min~I}$, $\epsilon_{\rm Min~II}$) that we added
quadratically to the published uncertainties. For the RVs of
\cite{Lacy:1998}, which have no published uncertainties, we assumed
errors of 2.5~\kms. The RV jitter terms were independent for the CfA
velocities and those of \cite{Lacy:1998}.

Initial solutions converged to a value of the radius ratio $k$ larger
than unity, which for main-sequence stars such as those in \vstar\ is
inconsistent with a derived mass ratio that is smaller than unity.
This is a common problem in fitting light curves of similar stars with
partial eclipses, and is due to strong correlations among several of
the parameters. In this case, external information must be used to
lift the degeneracies, such as a light ratio from spectroscopy
\citep[see, e.g.,][]{Andersen:1991}, which is very sensitive to the
radius ratio: $\ell_2/\ell_1 \propto k^2$. We therefore imposed a
Gaussian prior on the $V$-band light ratio generated at each step in
our Monte Carlo solution, set by the spectroscopic determination from
Section~\ref{sec:spectroscopy}, $\ell_2/\ell_1 = 0.92 \pm 0.02$. While
this measurement is not strictly in the $V$ passband, the difference
is negligible for our purposes because of the similarity in
temperature between the components.

\begin{deluxetable}{lc}
\tablewidth{0pt}
\tablecaption{Global solution for \vstar.\label{tab:results}}
\tablehead{
\colhead{~~~~~~~~~~~~Parameter~~~~~~~~~~~~} & \colhead{Value}
}
\startdata
$P_{\rm sid}$ (days)\dotfill               &  $15.33789922^{+0.00000057}_{-0.00000035}$\phn  \\ [+1ex]
$T_0$ (HJD-2,400,000)\dotfill              &  $54621.76727^{+0.00048}_{-0.00053}$\phm{2222}  \\ [+1ex]
$J(V)$\dotfill                             &  $0.9973^{+0.0063}_{-0.0044}$              \\ [+1ex]
$J({\rm KELT})$\dotfill                    &  $0.9984^{+0.0097}_{-0.0070}$                \\ [+1ex]
$r_1+r_2$\dotfill                          &  $0.08494^{+0.00042}_{-0.00042}$            \\ [+1ex]
$k \equiv r_2/r_1$\dotfill                 &  $0.9735^{+0.0073}_{-0.0096}$               \\ [+1ex]
$\cos i$\dotfill                           &  $0.00294^{+0.00034}_{-0.00056}$            \\ [+1ex]
$\sqrt{e}\cos\omega_0$\dotfill             &  $-0.08543^{+0.00021}_{-0.00021}$\phs       \\ [+1ex]
$\sqrt{e}\sin\omega_0$\dotfill             &  $-0.6790^{+0.0011}_{-0.0010}$\phs          \\ [+1ex]
$\dot\omega$ (deg century$^{-1}$)\dotfill  &  $0.859^{+0.042}_{-0.017}$                  \\ [+1ex]
$m_0(V)$\dotfill                           &  $0.7978^{+0.0016}_{-0.0015}$               \\ [+1ex]
$m_0({\rm KELT~east})$\dotfill             &  $14.91747^{+0.00042}_{-0.00036}$\phn       \\ [+1ex]
$m_0({\rm KELT~west})$\dotfill             &  $14.63733^{+0.00024}_{-0.00037}$\phn       \\ [+1ex]
$L_3(V)$\dotfill                           &  $0.0016^{+0.0082}_{-0.0016}$               \\ [+1ex]
$L_3({\rm KELT~east})$\dotfill             &  $0.1758^{+0.0096}_{-0.0090}$               \\ [+1ex]
$L_3({\rm KELT~west})$\dotfill             &  $0.1161^{+0.0071}_{-0.0104}$              \\ [+1ex]
$u(V)$\dotfill                             &  $0.339^{+0.063}_{-0.032}$                  \\ [+1ex]
$u({\rm KELT})$\dotfill                    &  $0.408^{+0.077}_{-0.153}$                  \\ [+1ex]
$f(V)$\dotfill                             &  $1.684^{+0.050}_{-0.055}$                  \\ [+1ex]
$f({\rm KELT~east})$\dotfill               &  $1.544^{+0.016}_{-0.020}$                  \\ [+1ex]
$f({\rm KELT~west})$\dotfill               &  $1.392^{+0.016}_{-0.021}$                  \\ [+1ex]
$\epsilon_{\rm Min~I}$ (days)\dotfill      &  $0.0019^{+0.0011}_{-0.0005}$               \\ [+1ex]
$\epsilon_{\rm Min~II}$ (days)\dotfill     &  $0.00054^{+0.00033}_{-0.00008}$            \\ [+1ex]
$\gamma_{\rm CfA}$ (\kms)\dotfill          &  $-15.48^{+0.22}_{-0.13}$\phn\phs           \\ [+1ex]
$\gamma_{\rm Lacy}$ (\kms)\dotfill         &  $-14.68^{+0.65}_{-1.04}$\phn\phs           \\ [+1ex]
$\Delta_{\rm CfA}$ (\kms)\dotfill          &  $-0.30^{+0.26}_{-0.24}$\phs                \\ [+1ex]
$\Delta_{\rm Lacy}$ (\kms)\dotfill         &  $+0.3^{+1.5}_{-1.3}$\phs                   \\ [+1ex]
$K_1$ (\kms)\dotfill                       &  $79.36^{+0.17}_{-0.27}$\phn                \\ [+1ex]
$K_2$ (\kms)\dotfill                       &  $81.97^{+0.20}_{-0.24}$\phn                \\ [+1ex]
$\epsilon_{\rm RV,1}$(CfA) (\kms)\dotfill  &  $0.06^{+0.40}_{-0.02}$                     \\ [+1ex]
$\epsilon_{\rm RV,2}$(CfA) (\kms)\dotfill  &  $0.10^{+0.71}_{-0.10}$                     \\ [+1ex]
$\epsilon_{\rm RV,1}$(Lacy) (\kms)\dotfill &  $2.5^{+1.0}_{-1.5}$                        \\ [+1ex]
$\epsilon_{\rm RV,2}$(Lacy) (\kms)\dotfill &  $3.1^{+1.3}_{-0.7}$                        \\ %[+0.5ex]
\noalign{\vskip 2pt}
\noalign{\hrule}
\noalign{\vskip 2pt}
\multicolumn{2}{c}{Derived quantities} \\
\noalign{\vskip 1pt}
\noalign{\hrule}
\noalign{\vskip 3pt}
$P_{\rm anom}$ (days)\dotfill              &  $15.33790024^{+0.00000056}_{-0.00000034}$\phn  \\ [+1ex]
$e$\dotfill                                &  $0.4684^{+0.0013}_{-0.0015}$               \\ [+1ex]
$\omega_0$ (deg)\dotfill                   &  $262.829^{+0.027}_{-0.030}$\phn\phn        \\ [+1ex]
$\dot\omega$ ($10^{-7}$ rad day$^{-1}$)\dotfill  & $4.11^{+0.20}_{-0.08}$                \\ [+1ex]
$U$ (years)\dotfill                        &  $41800^{+1000}_{-1800}$\phn                \\ [+1ex]
$i$ (deg)\dotfill                          &  $89.832^{+0.032}_{-0.019}$\phn             \\ [+1ex]
$r_1$\dotfill                              &  $0.04306^{+0.00024}_{-0.00023}$            \\ [+1ex]
$r_2$\dotfill                              &  $0.04188^{+0.00030}_{-0.00030}$            \\ [+1ex]
$q \equiv M_2/M_1$\dotfill                 &  $0.9681^{+0.0036}_{-0.0040}$               \\ [+1ex]
$a$ ($R_{\sun}$)\dotfill                   &  $43.198^{+0.093}_{-0.086}$\phn              \\ [+1ex]
$\ell_2/\ell_1(V)$\dotfill                 &  $0.942^{+0.016}_{-0.013}$                  \\ [+1ex]
$\ell_2/\ell_1({\rm KELT})$\dotfill        &  $0.944^{+0.017}_{-0.015}$                  \\ [+1ex]
Phase of Min~II at $T_0$\dotfill           &  $0.457997^{+0.000041}_{-0.000032}$
\enddata
\end{deluxetable}

The results of our global fit are presented in
Table~\ref{tab:results}, where we list for each parameter the mode of
the posterior distributions along with the 68.3\% credible
intervals. Additional properties derived from the fitted quantities
were computed by combining the corresponding Markov chains, link by
link, and are reported as well. As anticipated, third light is highly
significant for both of the KELT light curves, indicating an
appreciable amount of flux is coming from nearby field stars.
Contamination is larger to the east ($\sim$18\%) compared to the west
($\sim$12\%), which is also as expected. The fitted $L_3$ for the
$V$-band light curve of \cite{Khaliullin:1985} is formally below 0.2\%
and not significantly different from zero. Therefore, we find no
evidence for the $\Delta V \approx 1.2$~mag companion reported in the
{\it Tycho\/} Double Star Catalogue, which if real should contribute
about 25\% of the light in $V$. This star would be expected to be even
more prominent in the redder KELT passband, and yet the corresponding
$L_3$ values are smaller than 25\%. The radial-velocity offsets
$\Delta$ between the primary and secondary for both RV data sets are
also not statistically significant.  Repeating the fit with $L_3(V)$
and the velocity offsets set to zero leads to nearly identical values
for all elements. Nevertheless, we have chosen here to retain the
results that include these extra parameters, to be conservative, so
that any uncertainty in their determination is propagated through to
the rest of the adjusted quantities.

The linear limb-darkening coefficient for the $V$ band,
$0.339^{+0.063}_{-0.032}$, is somewhat lower than the value of $\sim$0.45
expected from theory for the mean temperature of the components
\citep[e.g.,][]{Claret:2011}. The coefficient obtained for the KELT
light curves, $0.408^{+0.077}_{-0.153}$, is close to the value of 0.38
predicted for the $R$ band but this is likely accidental as the
passbands are not exactly the same.

\begin{figure*}
\includegraphics[scale=0.73,angle=270]{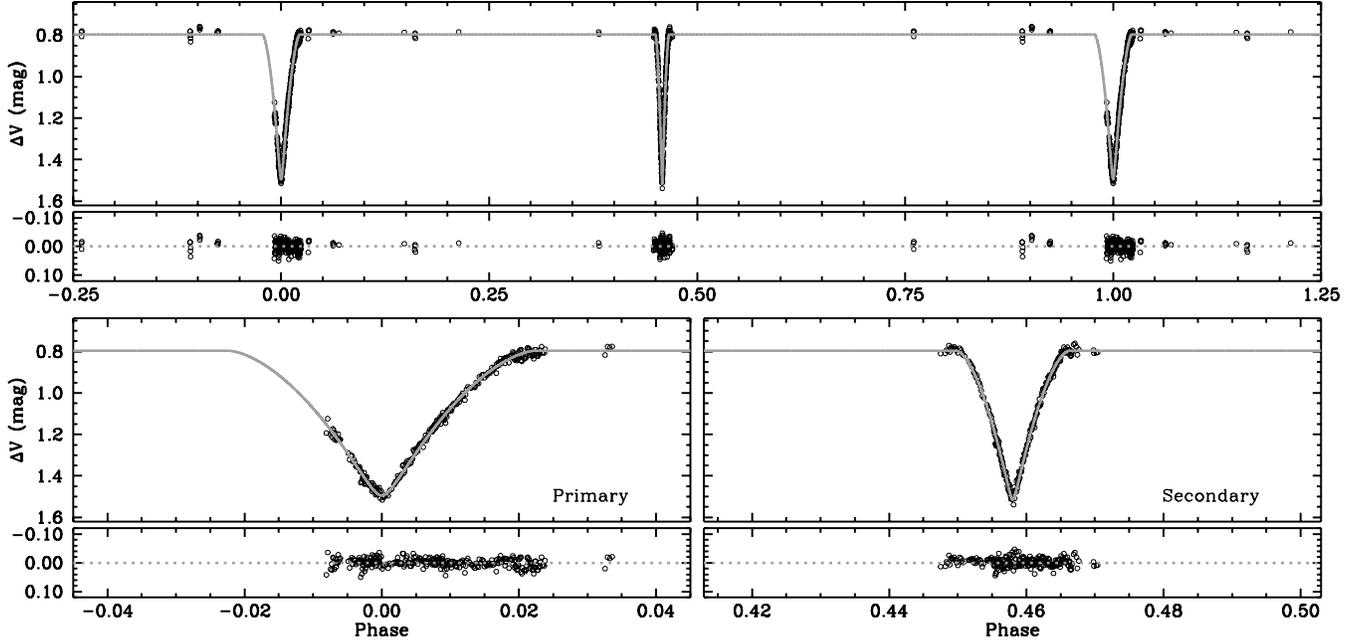}

\figcaption[]{Photometry from \cite{Khaliullin:1985} in the $V$ band
  along with our model light curve, reduced to the reference epoch
  $T_0$ in Table~\ref{tab:results}. Enlargements of the primary and
  secondary eclipses in the bottom panels have the same horizontal
  scale to permit a direct comparison of the widths of the
  eclipses. Residuals from the fit are shown in each panel.\label{fig:khaliullin}}
\end{figure*}

\begin{figure*}
\includegraphics[scale=0.73,angle=270]{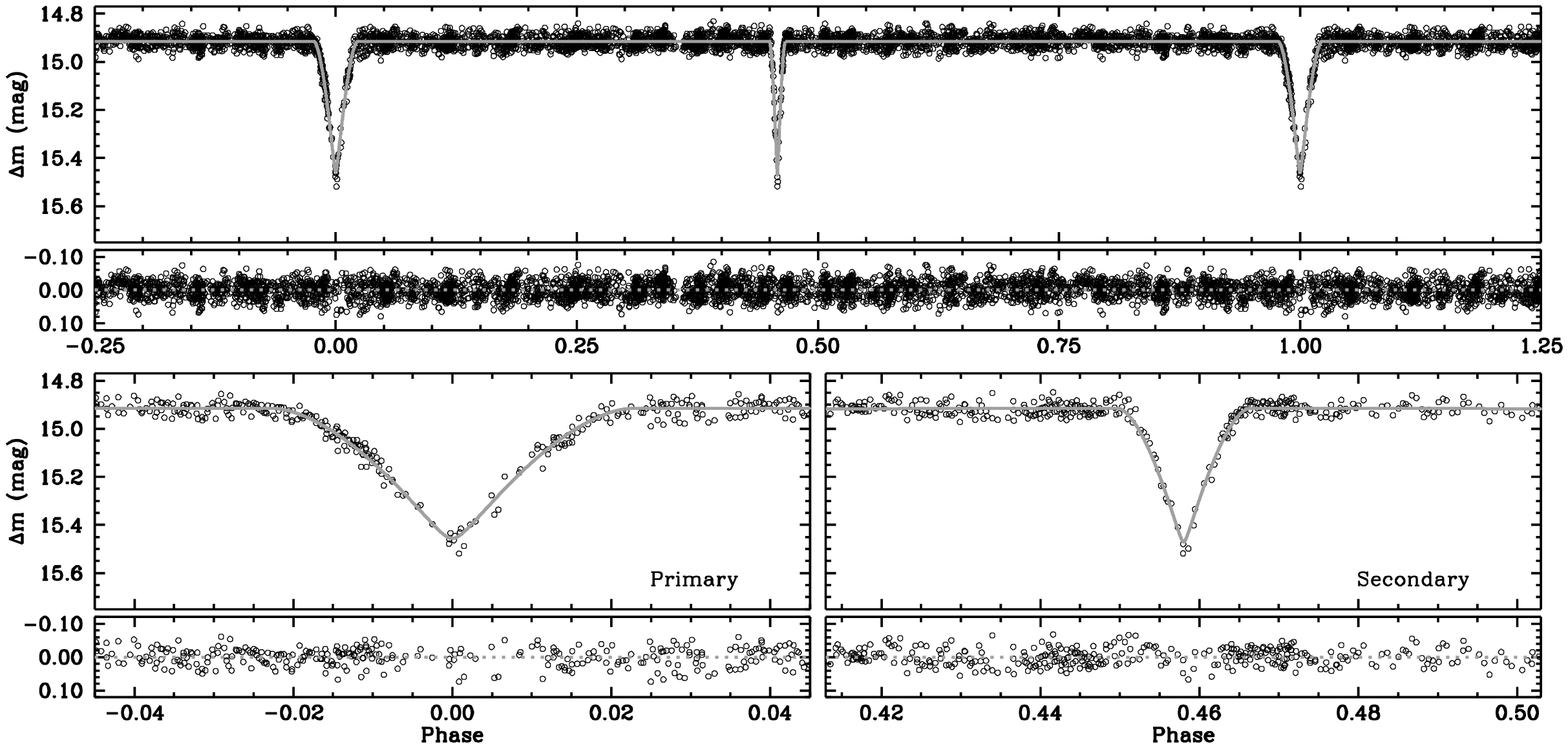}

\figcaption[]{Same as Figure~\ref{fig:khaliullin} for the KELT east
  photometry.\label{fig:KELTeast}}
\end{figure*}

\begin{figure*}
\includegraphics[scale=0.73,angle=270]{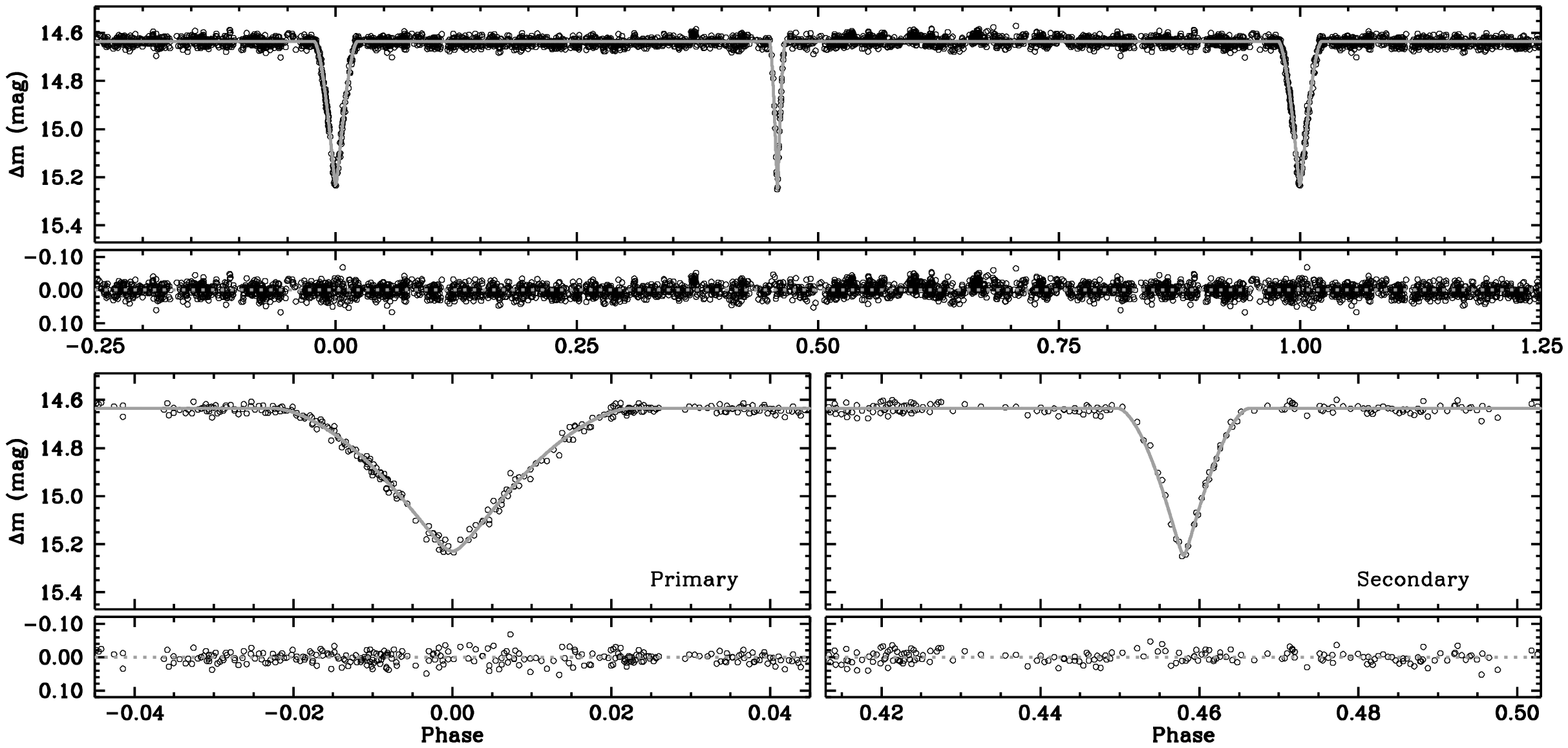}

\figcaption[]{Same as Figure~\ref{fig:khaliullin} for the KELT west
  photometry.\label{fig:KELTwest}}
\end{figure*}

\begin{figure}
\epsscale{1.15}
\plotone{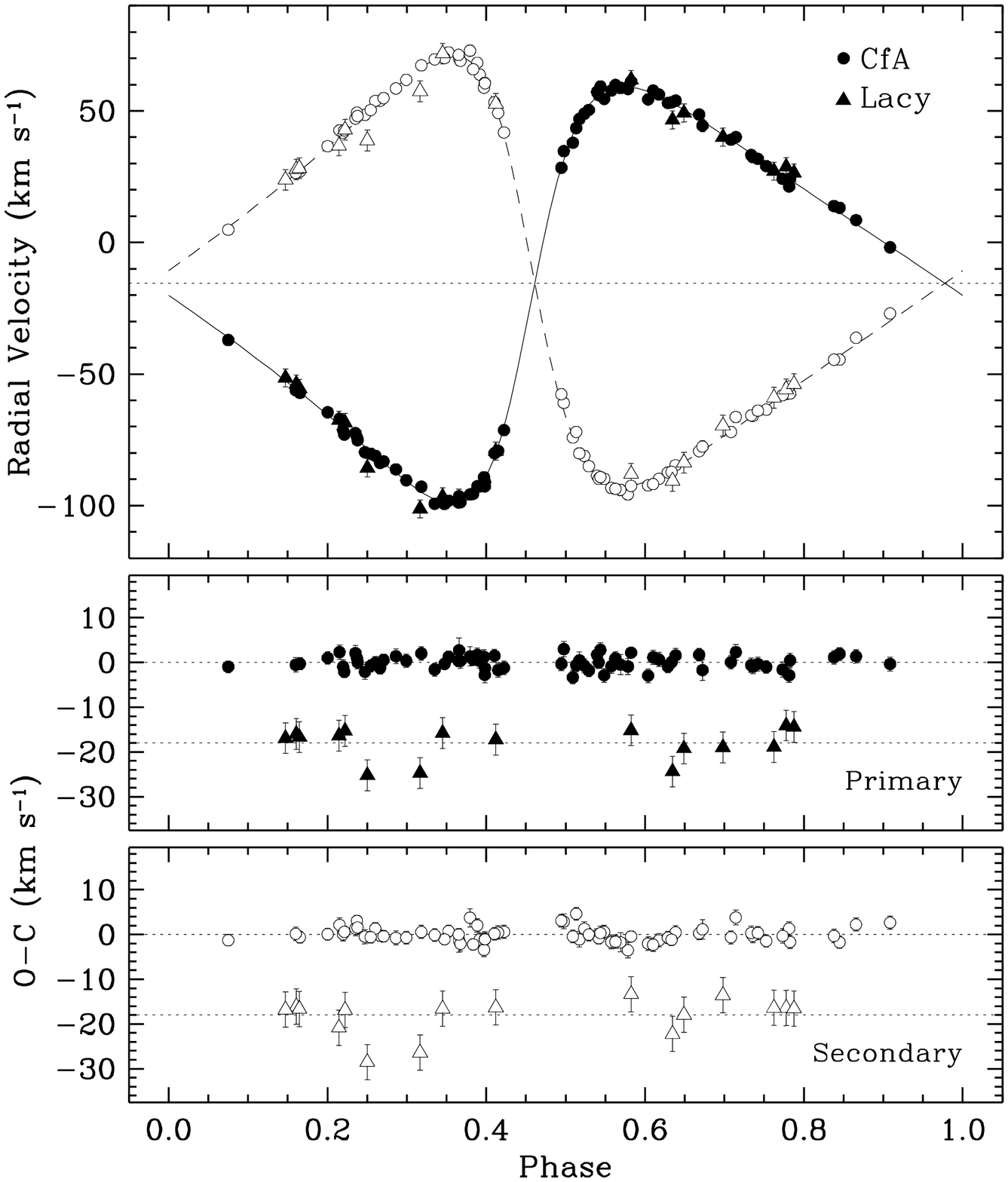}
\figcaption[]{Radial-velocity observations of \vstar\ and orbital
fit. Phase 0.0 corresponds to the time of primary eclipse, and the
dotted line is the center-of-mass velocity on the CfA reference
frame. Measurements for the primary are shown with filled
symbols. The observations by \cite{Lacy:1998} have been shifted
to the CfA frame for display purposes.\label{fig:rvs}}
\end{figure}

A graphical representation of the photometric observations and the
fitted light curves are given in Figure~\ref{fig:khaliullin} for the
$V$-band data of \cite{Khaliullin:1985}, and in
Figure~\ref{fig:KELTeast} and Figure~\ref{fig:KELTwest} for the KELT
measurements. In each case the residuals are shown below each
panel. The resulting scatter of the photometric measurements is
0.015~mag in $V$, 0.025~mag for KELT east, and 0.017~mag for KELT
west.  The CfA RV observations along with those of \cite{Lacy:1998}
are shown with the fitted model in Figure~\ref{fig:rvs}. For the CfA
data (71 measurements) the RMS residuals from the fit are
1.51~\kms\ and 1.63~\kms\ for the primary and secondary,
respectively. The 16 measurements by \cite{Lacy:1998} show a scatter
of 3.54~\kms\ and 4.10~\kms.

%%%%%%%%%%%%%%%%%%%%%%%%%%%%%%%%%%%%%%%%%%%%%%%%%%%%%%%%%%%%%%%%%%%%%%%%%%%
\section{Absolute dimensions}
\label{sec:dimensions}
%%%%%%%%%%%%%%%%%%%%%%%%%%%%%%%%%%%%%%%%%%%%%%%%%%%%%%%%%%%%%%%%%%%%%%%%%%%

The absolute masses and radii of \vstar\ are listed in
Table~\ref{tab:dimensions}\footnote{The physical constants used in
  this work conform to IAU Recommendation B3 \citep{Prsa:2016}.}.  The
relative uncertainties are smaller than 1\% in both properties and
represent a significant improvement over those obtained by
\cite{Lacy:1998}. Our temperature estimates from
Section~\ref{sec:spectroscopy} are marginally different for the two
components, and are somewhat higher than those obtained by
\cite{Guinan:1996}, $T_{\rm eff} = 9900 \pm 400$~K, and by
\cite{Lacy:1998}, $T_{\rm eff} = 9940 \pm 60$~K, both of whom assumed
identical temperatures for the two stars.

Str\"omgren photometry obtained by \cite{Lacy:2002} along with the
calibration by \cite{Crawford:1978} for B-type stars yields a mean
reddening of $E(b-y) = 0.066 \pm 0.005$, from which $(b-y)_0 = -0.027
\pm 0.005$ for the combined light. Tabulations by \cite{Popper:1980}
and \cite{Gray:2005} then yield mean temperatures of 10670~K and
10920~K, respectively.  Alternatively, the corresponding reddening in
the Johnson system, $E(B-V) = 0.089 \pm 0.007$, together with the
measured $B-V$ index of \cite{Lacy:1992} give $(B-V)_0 = -0.054 \pm
0.010$, from which we obtain mean temperatures of 10250~K
\citep{Popper:1980}, 10890~K \citep{Gray:2005}, and 10460~K
\citep{Pecaut:2013}. The average of the above five photometric
estimates, $10640 \pm 150$~K, is consistent with our spectroscopic
determinations (10650~K for the primary and 10350~K for the
secondary), supporting their accuracy. No spectroscopic metallicity is
available for \vstar; our own spectra are unsuitable for a detailed
analysis because of the narrow wavelength range, the relatively low
signal-to-noise ratios, and the degeneracy with temperature mentioned
earlier.

Table~\ref{tab:dimensions} lists our measured projected rotational
velocities for the components, which are considerably smaller than
earlier estimates of $20 \pm 5~\kms$ by \cite{Guinan:1996} and $24 \pm
2~\kms$ by \cite{Lacy:1998}. Also listed are the expected rotational
velocities assuming either pseudo-synchronous rotation
\citep{Hut:1981} or that the stars are synchronized to the mean
orbital motion. The measurements agree well with the
pseudo-synchronous values.

\begin{deluxetable}{lcc}
\tablewidth{0pt}
\tablecaption{Physical properties of \vstar.\label{tab:dimensions}}
\tablehead{
\colhead{~~~~~~~~~Parameter~~~~~~~~~} &
\colhead{Primary} &
\colhead{Secondary}
}
\startdata
Mass ($M_{\sun}$)\dotfill                          & $2.335^{+0.017}_{-0.013}$    &  $2.260^{+0.016}_{-0.013}$ \\ [+1ex]
Radius ($R_{\sun}$)\dotfill                        & $1.859^{+0.012}_{-0.009}$    &  $1.808^{+0.015}_{-0.013}$  \\ [+1ex]
$\log g$ (cgs)\dotfill                             & $4.2675^{+0.0052}_{-0.0050}$ &  $4.2770^{+0.0072}_{-0.0056}$  \\ [+0.5ex]
Temperature (K)\dotfill                            & 10650~$\pm$~200\phn\phn      &  10350~$\pm$~200\phn\phn \\ [+0.5ex]
$L/L_{\sun}$\dotfill                               & $39.9^{+3.3}_{-2.8}$\phn     &  $33.7^{+2.9}_{-2.5}$\phn  \\ [+0.5ex]
$BC_{\rm V}$ (mag)\tablenotemark{a}\dotfill        & $-$0.39~$\pm$~0.11\phs       &    $-$0.32~$\pm$~0.11\phs  \\ [+0.5ex]
$M_{\rm bol}$ (mag)\tablenotemark{b}\dotfill       & $0.721^{+0.086}_{-0.079}$    &  $0.906^{+0.089}_{-0.082}$ \\ [+0.5ex]
$M_V$ (mag)\dotfill                                & 1.11~$\pm$~0.14          &  1.23~$\pm$~0.14  \\ [+0.5ex]
$E(B-V)$ (mag)\dotfill                             & \multicolumn{2}{c}{0.089~$\pm$~0.007} \\ [+0.5ex]
$m-M$ (mag)\dotfill                                & \multicolumn{2}{c}{9.66~$\pm$~0.10} \\ [+0.5ex]
Distance (pc)\tablenotemark{c}\dotfill             & \multicolumn{2}{c}{$854^{+41}_{-38}$\phn} \\ [+0.5ex]
Parallax (mas)\dotfill                             & \multicolumn{2}{c}{$1.167^{+0.059}_{-0.050}$} \\ [+0.5ex]
$v_{\rm circ} \sin i$ (\kms)\dotfill               & $6.132^{+0.040}_{-0.031}$    &  $5.964^{+0.048}_{-0.042}$ \\ [+1ex]
$v_{\rm psync} \sin i$ (\kms)\dotfill              & $15.50^{+0.11}_{-0.13}$\phn  &  $15.09^{+0.09}_{-0.14}$\phn \\ [+0.5ex]
$v \sin i$ (\kms)\tablenotemark{d}\dotfill         &   15~$\pm$~1\phn             &    15~$\pm$~1\phn
\enddata
\tablenotetext{a}{Bolometric corrections from \cite{Flower:1996}, with
  a contribution of 0.10 mag added in quadrature to the uncertainty
  from the temperatures.}
\tablenotetext{b}{Uses $M_{\rm bol}^{\sun} = 4.732$ for consistency
  with the adopted table of bolometric
  corrections \citep[see][]{Torres:2010}.}
\tablenotetext{c}{Relies on the luminosities, the apparent
  magnitude of \vstar\ out of eclipse \citep[$V = 10.350 \pm
    0.008$;][]{Lacy:1992}, and bolometric corrections.}
\tablenotetext{d}{Measured value.}
\end{deluxetable}

%%%%%%%%%%%%%%%%%%%%%%%%%%%%%%%%%%%%%%%%%%%%%%%%%%%%%%%%%%%%%%%%%%%%%%%%%%%
\section{Comparison with stellar evolution models}
\label{sec:evolution}
%%%%%%%%%%%%%%%%%%%%%%%%%%%%%%%%%%%%%%%%%%%%%%%%%%%%%%%%%%%%%%%%%%%%%%%%%%%

\begin{figure}
\epsscale{1.15}
\plotone{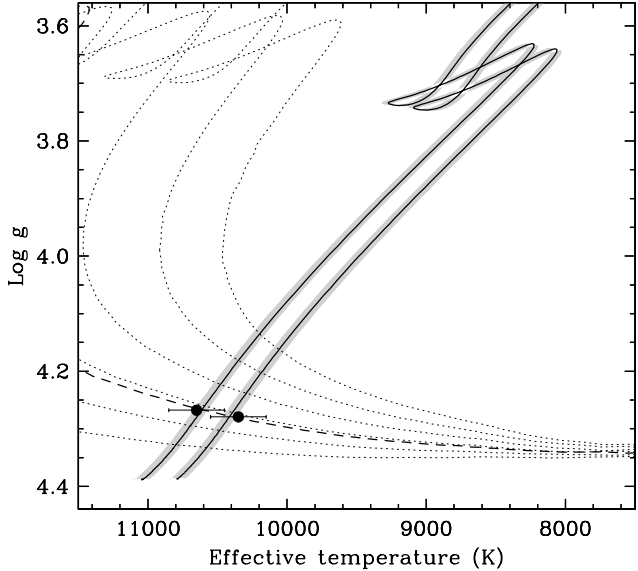}
\figcaption[]{Measurements for \vstar\ in the $\log g$ vs.\ $T_{\rm
    eff}$ diagram compared with evolutionary tracks from the MIST
  series \citep{Choi:2016} for a metallicity of ${\rm [Fe/H]} = -0.18$
  that best matches the observations. The shaded areas around the
  solid primary and secondary tracks represent the uncertainty in the
  measured masses. Isochrones ranging from 100 to 350~Myr are shown
  with dotted lines in steps of 50~Myr, with the best fit at 190~Myr
  indicated with a heavy dashed line.\label{fig:mistlogg}}
\end{figure}

\begin{figure}
\epsscale{1.15}
\plotone{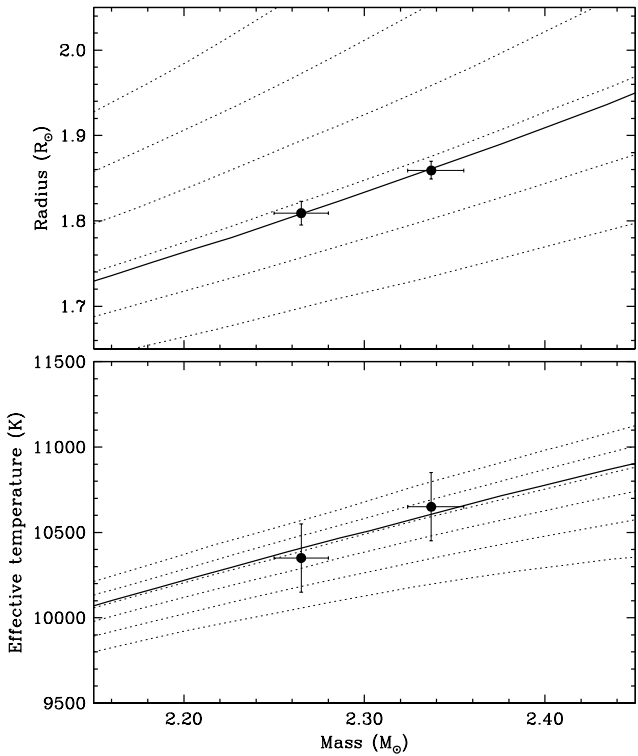}
\figcaption[]{Measured masses, radii, and temperatures of
  \vstar\ compared against model isochrones from the MIST series
  \citep{Choi:2016} for the same metallicity of ${\rm [Fe/H]} = -0.18$
  as in Figure~\ref{fig:mistlogg}. Dotted lines represent ages of 100--350~Myr
  in steps of 50~Myr, and the solid line is the best-fit age of
  190~Myr.\label{fig:mistmassradius}}
\end{figure}

The accurate properties derived for \vstar\ permit an interesting
comparison with predictions from current stellar evolution
theory. Figure~\ref{fig:mistlogg} shows these determinations in the
$\log g$ versus $T_{\rm eff}$ plane along with evolutionary tracks for
the measured masses from the MESA Isochrones and Stellar Tracks series
\citep[MIST;][]{Choi:2016} series, which is based on the Modules for
Experiments in Stellar Astrophysics package
\citep[MESA;][]{Paxton:2011, Paxton:2013, Paxton:2015}. The
metallicity in the models has been set to ${\rm [Fe/H]} = -0.18$
in order to match the observations. The models are in excellent
agreement with the observations at an age of 190~Myr; an isochrone for
this age is shown in the figure with a dashed line. In particular, the
temperature difference between the components determined from
spectroscopy is consistent with the separation between the
evolutionary tracks, which depends on the mass ratio. The stars are
seen to be little evolved from the zero-age main sequence.  The same
comparison is shown in the mass-radius and mass-temperature diagrams
of Figure~\ref{fig:mistmassradius}, also indicating good agreement.

%%%%%%%%%%%%%%%%%%%%%%%%%%%%%%%%%%%%%%%%%%%%%%%%%%%%%%%%%%%%%%%%%%%%%%%%%%%
\section{Apsidal motion}
\label{sec:apsidal}
%%%%%%%%%%%%%%%%%%%%%%%%%%%%%%%%%%%%%%%%%%%%%%%%%%%%%%%%%%%%%%%%%%%%%%%%%%%

Estimates of the rate of apsidal motion for \vstar\ have varied
considerably over the years, beginning with the first measurement by
\citep{Khaliullin:1983} yielding $\dot\omega = 1.04 \pm
0.15$~deg~century$^{-1}$, later revised to $0.90 \pm
0.13$~deg~century$^{-1}$ \citep{Khaliullin:1985}. A similar estimate
of $0.95 \pm 0.06$~deg~century$^{-1}$ was published by
\cite{Lines:1989}.  All three studies judged these values to be
consistent with expectations from theory (including classical terms
and General Relativity), although the physical parameters of the
components necessary to compute the predicted apsidal motion rate were
not particularly well known at the time, as no dynamical masses were
available.  Much smaller values of $\dot\omega$ were reported
subsequently by \cite{Wolf:1995}, $0.53 \pm 0.11$~deg~century$^{-1}$,
\cite{Guinan:1996}, $0.52 \pm 0.14$~deg~century$^{-1}$, and
\cite{Lacy:1998}, $0.60 \pm 0.10$~deg~century$^{-1}$, all of whom
concluded that the apsidal motion of \vstar\ was too slow compared to
theory. Finally, \cite{Volkov:1999} and \cite{Wolf:2010} reported
intermediate values of $0.86 \pm 0.05$~deg~century$^{-1}$ and $0.76
\pm 0.14$~deg~century$^{-1}$, respectively, and considered these to be
in good agreement with theory.

\begin{figure}
\epsscale{1.15}
\plotone{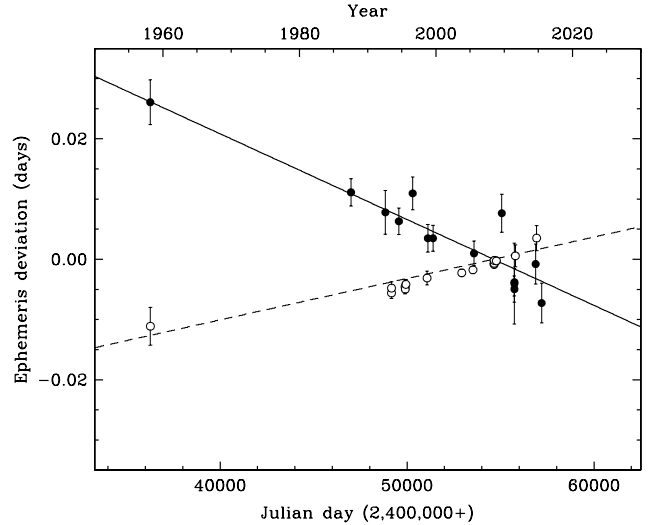}
\figcaption[]{Measured times of minimum light from
  Table~\ref{tab:timings}, and the ephemeris curve from our global
  fit. Filled circles correspond to primary
  timings..\label{fig:ephemeris}}
\end{figure}

The value resulting from our global fit (Table~\ref{tab:results}),
$\dot\omega = 0.859^{+0.042}_{-0.017}$~deg~century$^{-1}$, is more in
line with the recent studies and is considerably more precise. It
corresponds to an apsidal period of $U = 41800^{+1000}_{-1800}$
years. The measurements and the computed ephemeris curve are shown in
Figure~\ref{fig:ephemeris}.

With the accurate knowledge we now have of the physical properties of
the components, we estimate the predicted rate of periastron advance
from classical terms (tidal and rotational distortions) to be
$\dot\omega_{\rm clas} =
0.0988^{+0.0062}_{-0.0054}$~deg~century$^{-1}$, where we have adopted
identical internal structure constants for the two stars of $\log k_2
= -2.33 \pm 0.03$ from the models by \cite{Claret:2004}, for the
average mass and age of the system. The general relativistic
contribution \citep[e.g.,][]{Levi-Civita:1937, Gimenez:1985} is
calculated to be $\dot\omega_{\rm GR} =
0.7447^{+0.0032}_{-0.0031}$~deg~century$^{-1}$, which is 7.5 times
larger than the classical effect. The total expected apsidal motion is
then $\dot\omega_{\rm tot} =
0.8435^{+0.0073}_{-0.0065}$~deg~century$^{-1}$. This agrees with our
measurement within the uncertainties (at the 0.8$\sigma$ level).

%%%%%%%%%%%%%%%%%%%%%%%%%%%%%%%%%%%%%%%%%%%%%%%%%%%%%%%%%%%%%%%%%%%%%%%%%%%
\section{Discussion and final remarks}
\label{sec:discussion}
%%%%%%%%%%%%%%%%%%%%%%%%%%%%%%%%%%%%%%%%%%%%%%%%%%%%%%%%%%%%%%%%%%%%%%%%%%%

Our measurements of the properties of \vstar\ represent an improvement
in the precision of the masses of a factor of $\sim$5 over the work of
\cite{Lacy:1998}, and a factor of $\sim$3 in the radii. Much of this
is due to the greater number and higher resolution of our
spectroscopic observations, and partly also to the additional light
curves from KELT. The absolute dimensions of the system now rank among
the best for eclipsing binaries \citep[see, e.g.,][]{Torresetal:2010}.

The agreement with stellar evolution theory is excellent, and suggests
an age of 190~Myr according to the MIST models. The abundance we infer
from this comparison, ${\rm [Fe/H]} = -0.18$, is not out of the
ordinary. The $UVW$ space motion components based on our
center-of-mass velocity (CfA zero point), distance, and the proper
motion from the first data release (DR1) of the {\it Gaia\/} mission
\citep{Lindegren:2016} are $U = +1.6 $~\kms, $V = -6.8$~\kms, and $W =
-9.3$~\kms\ in the LSR frame.\footnote{$U$ is counted positive toward
  the Galactic center.}  These are typical of the thin disk, as
expected from the youth of the system.

\vstar\ is a member of a very small group of eclipsing systems with
well determined properties and accurately measured apsidal motion in
which the contribution from General Relativity is significant. In this
particular example the GR effect represents 88\% of the total
predicted motion.  Estimates of $\dot\omega$ over the last 30 years
have relied almost exclusively on measurements of times of minimum
light, and have varied by nearly a factor of two depending largely on
the sample of timings used. In addition to having an updated list of
historical timings going back to 1958, in this work we have added to
the analysis radial velocities spanning 22 years and light curves
separated in time by about 30 years, all of which provide additional
information constraining the rate of periastron advance. As a result,
the formal uncertainty of $\dot\omega$ has been reduced significantly
to better than 5\%.  We find that the observed rate agrees with the
predicted rate within the errors, consistent with some of the more
recent but less precise estimates from the last 15 years. Even so, it
would be of considerable interest to attempt a determination of the
orientation of the spin axes of the \vstar\ components through the
Rossiter-McLaughlin effect, which was used successfully in DI~Her to
show that the axes of those stars are tilted relative to the orbit,
explaining the anomalously slow apsidal motion of that system that had
puzzled astronomers for decades. Since there does not appear to be any
disagreement with the expected value of $\dot\omega$ for \vstar, we
would not expect the spin axes to be strongly misaligned.

Our accurate distance estimate for \vstar\ leads to an inferred
parallax of $1.167^{+0.059}_{-0.050}$~mas that is marginally
consistent with the trigonometric value of $0.70 \pm 0.34$~mas from
the first data release (DR1) of the {\it Gaia\/} mission, where the
uncertainty quoted in the last value does not include an estimated
contribution of 0.30~mas from systematic errors
\citep{Lindegren:2016}. Recent studies \citep{Stassun:2016, Jao:2016}
have indicated there may in fact be a small but significant bias in
the {\it Gaia\/} parallaxes of about 0.25~mas, in the sense that the
original {\it Gaia\/} values are too small. We note that this happens
to go in the direction of bringing agreement in the case of \vstar.

Finally, a detailed abundance analysis of \vstar\ would be highly
beneficial as it would permit a more stringent comparison with stellar
evolution models than performed here. Because a change in abundance
shifts the evolutionary tracks horizontally in
Figure~\ref{fig:mistlogg}, knowing the metallicity could serve to test
the accuracy of the temperature estimates from this work.

\acknowledgments

We are grateful to P.\ Berlind, M.\ Calkins, R.\ J.\ Davis,
D.\ W.\ Latham, and R.\ P.\ Stefanik for help in obtaining the
spectroscopic observations of \vstar, and to R.\ J.\ Davis and
J.\ Mink for maintaining the CfA echelle database over the years. We
also thank J.\ Irwin for helpful discussions about the use of his
light-curve code and for implementing the apsidal motion capability at
our request, A.\ Claret for providing the stellar evolution tracks
used for the apsidal motion calculation, and the anonymous referee for
a helpful comment. G.T.\ acknowledges partial support for this work
from NSF grant AST-1509375.  The efforts of C.M.\ were supported by
the SAO REU program, funded in part by the National Science Foundation
REU and Department of Defense ASSURE programs under NSF Grant
No.\ 1262851, and by the Smithsonian Institution. Work performed by
J.E.R.\ was supported by the Harvard Future Faculty Leaders
Postdoctoral fellowship. This research has made use of the SIMBAD and
VizieR databases, operated at CDS, Strasbourg, France, and of NASA's
Astrophysics Data System Abstract Service.

%%%%%%%%%%%%%%%%%%%%%%%%%%%%%%%%%%%%%%%%%%%%%%%%%%%%%%%%%%%%%%%%%%%%%%%%%%%

\end{document}